\documentclass[fleqn,usenatbib]{mnras}

\usepackage{newtxtext,newtxmath}

\usepackage[T1]{fontenc}
\usepackage{ae,aecompl}

\usepackage{graphicx}	
\usepackage{amsmath}	

\title[ASAS-SN Catalog of Variable Stars III]{The ASAS-SN Catalog of Variable Stars III: \textit{Variables in the Southern TESS Continuous Viewing Zone}}

\author[T. Jayasinghe et al.]{T. Jayasinghe$^{1,2}$\thanks{E-mail: jayasinghearachchilage.1@osu.edu},
K. Z. Stanek$^{1,2}$,
C. S. Kochanek$^{1,2}$,
B. J. Shappee$^{3}$,
\newauthor 
T. W. -S. Holoien$^{4}$,
Todd A. Thompson$^{1,2,5}$,
J. L. Prieto$^{6,7}$,
Subo Dong$^{8}$,
M. Pawlak$^{9}$,
\newauthor 
O. Pejcha$^{9}$,
J. V. Shields$^{1}$,
G. Pojmanski$^{10}$,
S. Otero$^{11}$,
N. Hurst$^{12}$,
C. A. Britt$^{12}$,
\newauthor 
D. Will$^{1,12}$
\\
$^{1}$Department of Astronomy, The Ohio State University, 140 West 18th Avenue, Columbus, OH 43210, USA\\
$^{2}$Center for Cosmology and Astroparticle Physics, The Ohio State University, 191 W. Woodruff Avenue, Columbus, OH 43210, USA\\
$^{3}$Institute for Astronomy, University of Hawaii, 2680 Woodlawn Drive, Honolulu, HI 96822,USA\\
$^{4}$Carnegie Observatories, 813 Santa Barbara Street, Pasadena, CA 91101, USA\\
$^{5}$Institute for Advanced Study, Princeton, NJ, 08540\\
$^{6}$N\'ucleo de Astronom\'ia de la Facultad de Ingenier\'ia y Ciencias, Universidad Diego Portales, Av. Ej\'ercito 441, Santiago, Chile\\
$^{7}$Millennium Institute of Astrophysics, Santiago, Chile\\
$^{8}$Kavli Institute for Astronomy and Astrophysics, Peking University, Yi He Yuan Road 5, Hai Dian District, China\\
$^{9}$Institute of Theoretical Physics, Faculty of Mathematics and Physics, Charles University in Prague, Czech Republic\\
$^{10}$Warsaw University Observatory, Al Ujazdowskie 4, 00-478 Warsaw, Poland\\
$^{11}$The American Association of Variable Star Observers, 49 Bay State Road, Cambridge, MA 02138, USA\\
$^{12}$ASC Technology Services, 433 Mendenhall Laboratory 125 South Oval Mall Columbus OH, 43210, USA\\
}

\date{Accepted XXX. Received YYY; in original form ZZZ}

\pubyear{2018}

\begin{document}
\label{firstpage}
\pagerange{\pageref{firstpage}--\pageref{lastpage}}
\maketitle

\begin{abstract}
The All-Sky Automated Survey for Supernovae (ASAS-SN) provides long baseline (${\sim}4$ yrs) light curves for sources brighter than V$\lesssim17$ mag across the whole sky. The Transiting Exoplanet Survey Satellite (TESS) has started to produce high-quality light curves with a baseline of at least 27 days, eventually for most of the sky. The combination of ASAS-SN and TESS light curves probes both long and short term variability in great detail, especially towards the TESS continuous viewing zones (CVZ) at the ecliptic poles. We have produced ${\sim}1.3$ million V-band light curves covering a total of ${\sim}1000 \, \rm deg^2$ towards the southern TESS CVZ and have systematically searched these sources for variability. We have identified ${\sim} 11,700$ variables, including ${\sim} 7,000$ new discoveries. The light curves and characteristics of the variables are all available through the ASAS-SN variable stars database (\url{https://asas-sn.osu.edu/variables}). We also introduce an online resource to obtain pre-computed ASAS-SN V-band light curves (\url{https://asas-sn.osu.edu/photometry}) starting with the light curves of the ${\sim}1.3$ million sources studied in this work. This effort will be extended to provide ASAS-SN light curves for ${\sim}50\;$million sources over the entire sky. 

\end{abstract}




\section{Introduction}
The study of stellar variability has been invigorated by the advent of modern large scale sky surveys in the modern era. Recent surveys such as the All-Sky Automated Survey (ASAS; \citealt{2002AcA....52..397P}), the Optical Gravitational Lensing Experiment (OGLE; \citealt{2003AcA....53..291U}),the Northern Sky Variability Survey (NSVS; \citealt{2004AJ....127.2436W}), MACHO \citep{1997ApJ...486..697A}, EROS \citep{2002A&A...389..149D}, the Catalina Real-Time Transient Survey (CRTS; \citealt{2014ApJS..213....9D}), the Asteroid Terrestrial-impact Last Alert System (ATLAS; \citealt{2018PASP..130f4505T,2018arXiv180402132H}), and Gaia \citep{2018arXiv180409365G,2018arXiv180409373H,gdr2var} have collectively discovered $\gtrsim 10^6$ variables.

Variable stars are excellent astrophysical probes and have been used in numerous astronomical contexts. Pulsating variables such as Cepheids and RR Lyrae stars are commonly used as distance indicators owing to the period luminosity relationships seen amongst these variables (e.g., \citealt{1908AnHar..60...87L,2006MNRAS.370.1979M,2018SSRv..214..113B}, and references therein). Eclipsing binary stars are excellent probes of stellar systems and with sufficient radial velocity followup, allow for the derivation of useful astrophysical parameters, including the masses and radii, of the stars in these systems \citep{2010A&ARv..18...67T}. Variable stars are also useful for the study of stellar populations and Galactic structure \citep{2018IAUS..334...57M,2014IAUS..298...40F} .

Until recently, the All-Sky Automated Survey for SuperNovae (ASAS-SN, \citealt{2014ApJ...788...48S, 2017PASP..129j4502K}) monitored the visible sky to a depth of $V\lesssim17$ mag with a cadence of 2-3 days using two units in Chile and Hawaii each with 4 telescopes. Starting in 2017, ASAS-SN expanded to 5 units with 20 telescopes. The 3 new units all started with g-band filters and the 2 original units have now switched to g-band as well. The ASAS-SN telescopes are hosted by the Las Cumbres Observatory (LCO; \citealt{2013PASP..125.1031B}) in Hawaii, Chile, Texas and South Africa. ASAS-SN primarily focuses on the detection of bright supernovae (e.g., \citealt{2017MNRAS.471.4966H,2018arXiv181108904H}), tidal disruption events (e.g.,\citealt{2014MNRAS.445.3263H,2016MNRAS.455.2918H,2018arXiv180802890H}) and other transients (e.g.,  \citealt{2018arXiv180807875T,2018RNAAS...2b...8R}), but its excellent baseline allows for the study of variability amongst the $\gtrsim50$ million bright ($V<17$ mag) sources across the whole sky. ASAS-SN team members have also studied the relative specific Type Ia supenovae rates \citep{2018arXiv181000011B} and the largest amplitude M-dwarf flares seen in ASAS-SN \citep{2018arXiv180904510S}.
 
In Paper I \citep{2018MNRAS.477.3145J}, we reported ${\sim}66,000$ new variables that were flagged during the search for supernovae, most of which are located in regions close to the Galactic plane or Celestial poles which were not well-sampled by previous surveys. In Paper II \citep{2018arXiv180907329J}, we uniformly analyzed ${\sim} 412,000$ known variables from the VSX catalog, and developed a robust variability classifier utilizing the ASAS-SN V-band light curves and data from external catalogues. We have also explored the synergy between ASAS-SN and APOGEE \citep{2015AJ....150..148H} with the discovery of the first likely non-interacting binary composed of a black hole with a field red giant \citep{2018arXiv180602751T} and a detailed variability analysis of the APOGEE sources to identify 1914 periodic variables (Pawlak et al., in prep). We have also identified rare variables, including 2 very long period detached eclipsing binaries \citep{2018RNAAS...2c.181J,2018RNAAS...2c.125J} and 19 R Coronae Borealis stars \citep{2018arXiv180904075S} .

The Transiting Exoplanet Survey Satellite (TESS; \citealt{2015JATIS...1a4003R}) will produce a large number of high-quality light curves with a baseline of at least 27 days for most of the sky. The TESS input catalog (TIC; \citealt{2018AJ....156..102S}) contains ${\sim} 470$ million sources, out of which $200,000$ selected targets are observed at a 2 min cadence, while the remaining sources are observed with a cadence of 30 min. \citet{2018AJ....155...39O} recently identified variable sources in a sample of $4$ million TIC sources, but did not classify these variables into explicit types. These sources were classified in paper II using ASAS-SN data.

Sources closer to the TESS continous viewing zone (CVZ) will be observed for a substantially longer period, approaching one year and ${\sim}15,000$ epochs at the ecliptic poles. These TESS light curves will probe short period variability in great detail. ASAS-SN provides long baseline (${\gtrsim}4$ yr) light curves sampled at a cadence of ${\sim} 1-3$ days, that complement the TESS light curves. 

We extracted the ASAS-SN light curves of ${\sim}1.3$ million sources within 18 deg of the Southern Ecliptic Pole. These sources are within the Southern TESS CVZ and will have well-sampled TESS light curves. In this work, we systematically search this sample for variable sources. This is, in part, a test run for carrying out such a search over the entire sky. In Section $\S2$, we discuss the ASAS-SN observations and data reduction procedures. Section $\S3$ discusses the variability search and classification procedures. In Section $\S4$, we discuss our results and present a summary of our work in Section $\S5$. All the light curves of these sources are made available to the public through our online database.

\section{Observations and Data reduction}
\label{data}
We started with the AAVSO Photometric All-Sky Survey (APASS; \citealt{2015AAS...22533616H}) DR9 catalog as our input source catalog. We selected all the APASS sources with $V<17$ mag in all the ASAS-SN fields with central field coordinates within 18 deg from the Southern Ecliptic Pole ($\alpha=90$ deg, $\delta=-66.55$ deg). This resulted in a list of ${\sim}1.3$M sources spanning a total of ${\sim}1,000 \, \rm deg^2$. ASAS-SN V-band observations were made by the ``Brutus" (Haleakala, Hawaii) and ``Cassius" (CTIO, Chile) quadruple telescopes between 2013 and 2018. Each ASAS-SN field has ${\sim}$ 200-600 epochs of observation to a depth of $V\lesssim17$ mag. Each camera has a field of view of 4.5 deg$^2$, the pixel scale is 8\farcs0 and the FWHM is ${\sim}$ 2 pixels. ASAS-SN nominally saturates at ${\sim} 10-11$ mag, but light curves of saturated sources are sometimes quite good due to corrections made for bleed trails (see \citealt{2017PASP..129j4502K}).

The light curves for these sources were extracted as described in \citet{2018MNRAS.477.3145J} using image subtraction \citep{1998ApJ...503..325A,2000A&AS..144..363A} and aperture photometry on the subtracted images with a 2 pixel radius aperture. The APASS catalog was used for calibration. The zero point offsets between the different cameras were corrected as described in \citet{2018MNRAS.477.3145J}.

\begin{figure}
	\includegraphics[width=0.5\textwidth]{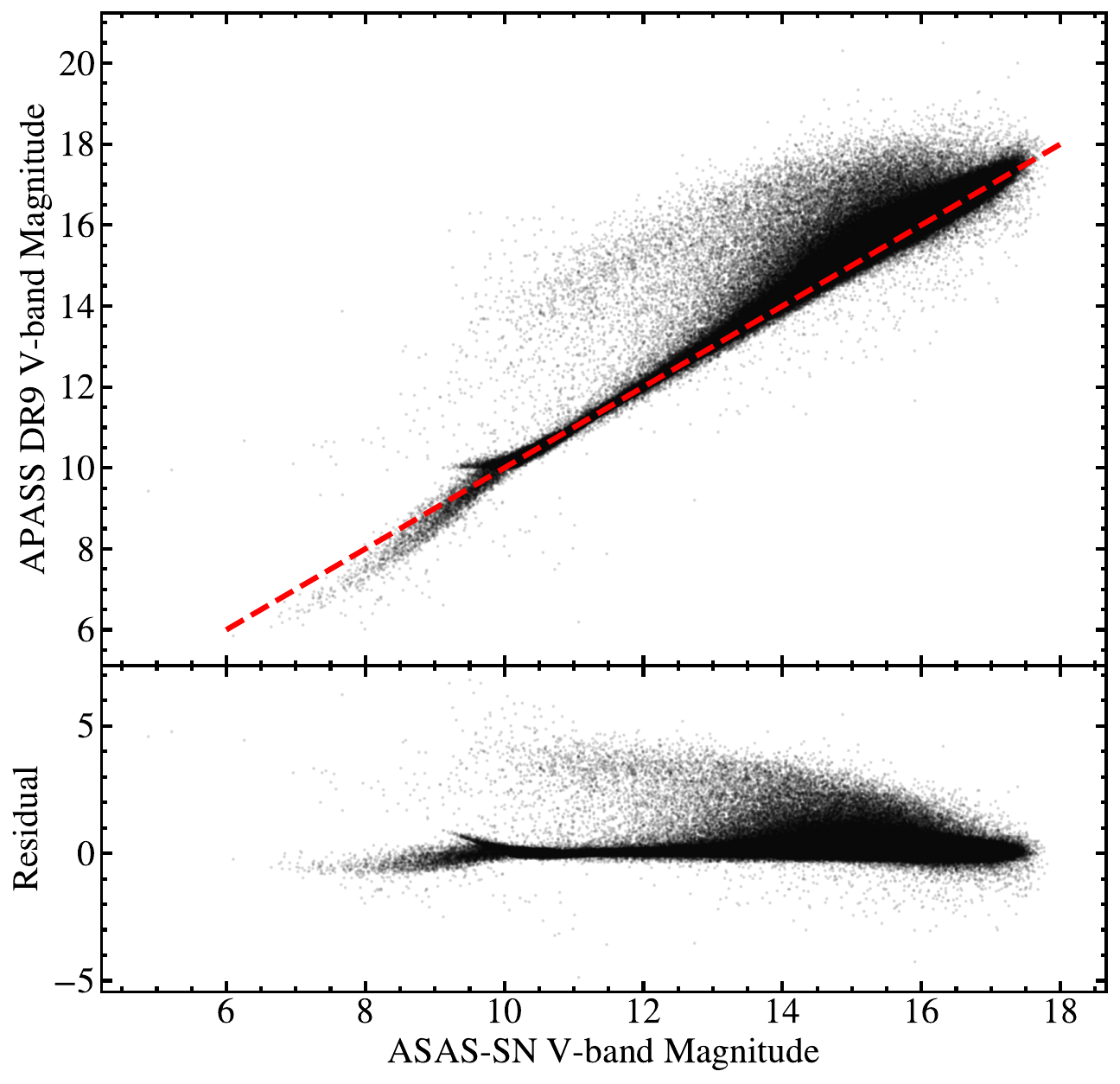}
    \caption{Comparison of the mean ASAS-SN V-band magnitudes to the APASS DR9 V-band magnitudes. The red dashed line illustrates a perfect calibration.}
    \label{fig:fig1}
\end{figure}

Figure \ref{fig:fig1} illustrates the relationship between the measured mean ASAS-SN V-band magnitudes and the APASS DR9 V-band magnitudes. Sources with $V_{\rm mean}\lesssim14$ mag have similar V-band magnitudes, but a large fraction of the sources with $V_{\rm mean}\gtrsim14$ mag show a discrepancy between the ASAS-SN and APASS measurements. This is due to blending in fields with significant stellar densities (i.e., the LMC in this work). The relatively large ASAS-SN pixel scale of 8\farcs0 as compared to the smaller APASS pixel scale of 2\farcs6 makes ASAS-SN photometry more susceptible to blending. Thus, ASAS-SN V-band measurements are systematically brighter for most sources in the LMC due to blended light.

The light curve extraction provides a statistical error estimate, but the scatter in the light curves of apparently non-variable sources is generally larger than expected given the nominal statistical uncertainties. In \citet{2018MNRAS.477.3145J}, we used the reduced $\chi^2$ statistic, \begin{equation}
    \frac{\chi^2}{N_{\rm DOF}}=\frac{1}{N-1}\sum_{i=1}^{N} \left(\frac{V_i -V_{\rm mean}}{\sigma(V_{\rm mean})}\right)^2\approx 1\,,
	\label{eq:chi2}
\end{equation} where $V_i$ is the V-band magnitude for epoch $i$, and $V_{\rm mean}$ is the mean V-band magnitude to determine the typical total uncertainty $\sigma(V_{\rm mean})$ as a function of magnitude. In this work, we update our approach to estimating the systematic uncertainties in the ASAS-SN photometry.

\begin{figure}
	\includegraphics[width=0.5\textwidth]{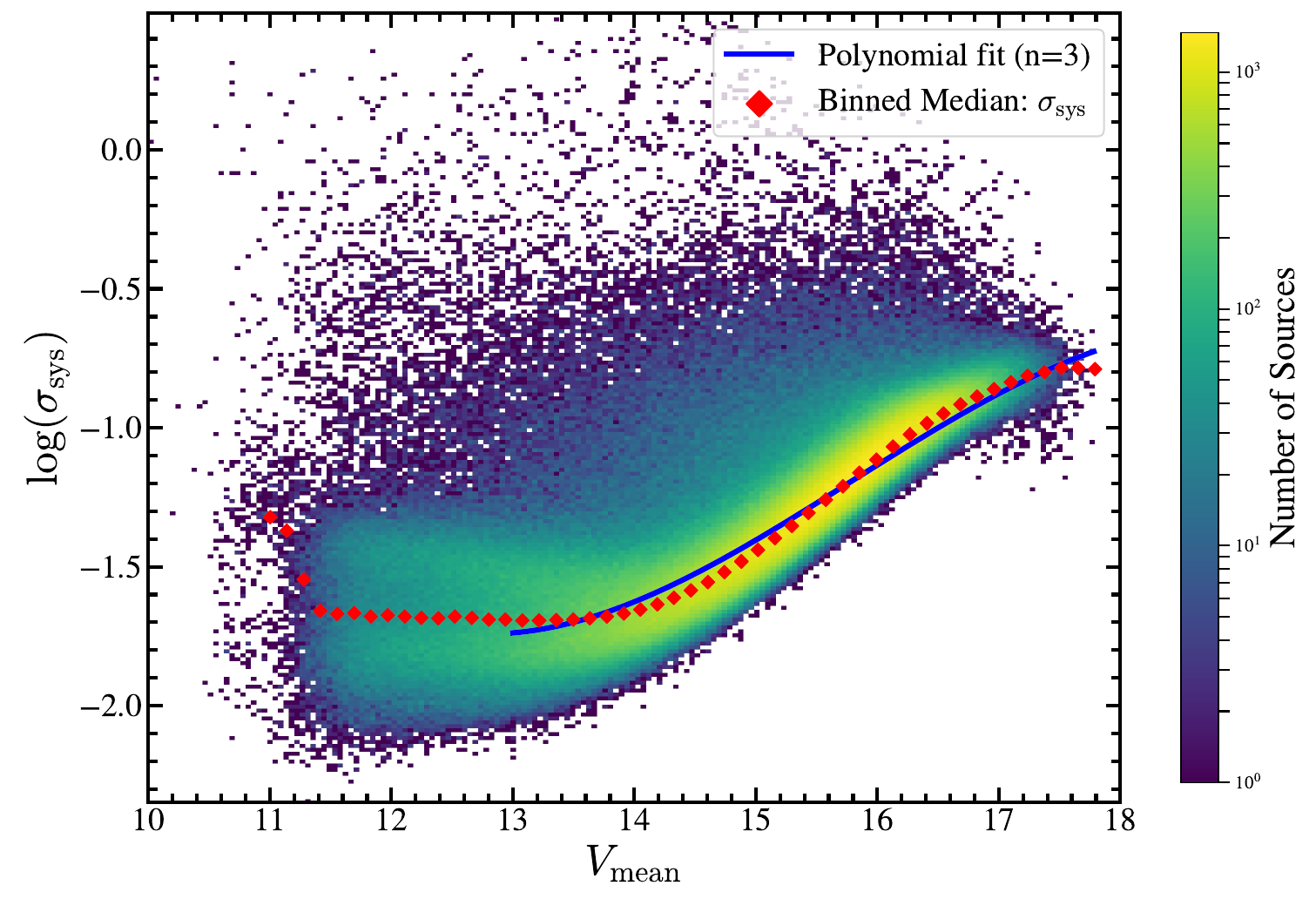}
    \caption{The distribution of $\sigma_{\rm sys}$ for the ${\sim}1.3$M sources. The red points are the binned median values of $\sigma_{\rm sys}$ for this sample in bins of $\sim0.25$ mag. The blue curve shows a third order polynomial model for sources with $V>13$ mag (Equation \ref{eq:errfix}).}
    \label{fig:fig2}
\end{figure}

We can view the photometric errors as the quadrature sum of the estimated statistical uncertainty $\sigma_{\rm stat}$ and a systematic $\sigma_{\rm sys}$ , with \begin{equation}
    \sigma^2=\sigma_{\rm sys}^2+\sigma_{\rm stat}^2\,.
	\label{eq:err}
\end{equation} We can measure $\sigma$ as the root-mean-square (rms) scatter in the light curves of non-variable stars, and then subtract the estimated statistical errors to derive $\sigma_{\rm sys}$, with the results shown in Figure \ref{fig:fig2} for the $1.3$M sources in our sample.  For sources with $V_{\rm mean}<13$ mag, the photometric uncertainties approach an ASAS-SN error floor of ${\sim}0.02$ mag. For the sources with $V_{\rm mean}>13$ mag, we fit a third order polynomial, \begin{equation}
	\log \sigma_{\rm sys}=A(x-13)^3+B(x-13)^2+C(x-13)-1.738\,,
	\label{eq:errfix}
\end{equation} with $A=-1.03\times 10^{-2}$, $B=8.6\times 10^{-2}$, and $C=3.6\times 10^{-2}$ to the median of the  $\sigma_{\rm sys}$ versus $V_{\rm mean}$ distribution. The polynomial smoothly joins to 0.02 mag at $V=13$ mag. To correct the errors in the light curves, we replace the formal magnitude errors by using equation \ref{eq:err} and either equation \ref{eq:errfix} or 0.02 mag for $\sigma_{\rm sys}$.

\begin{figure}
	\includegraphics[width=0.5\textwidth]{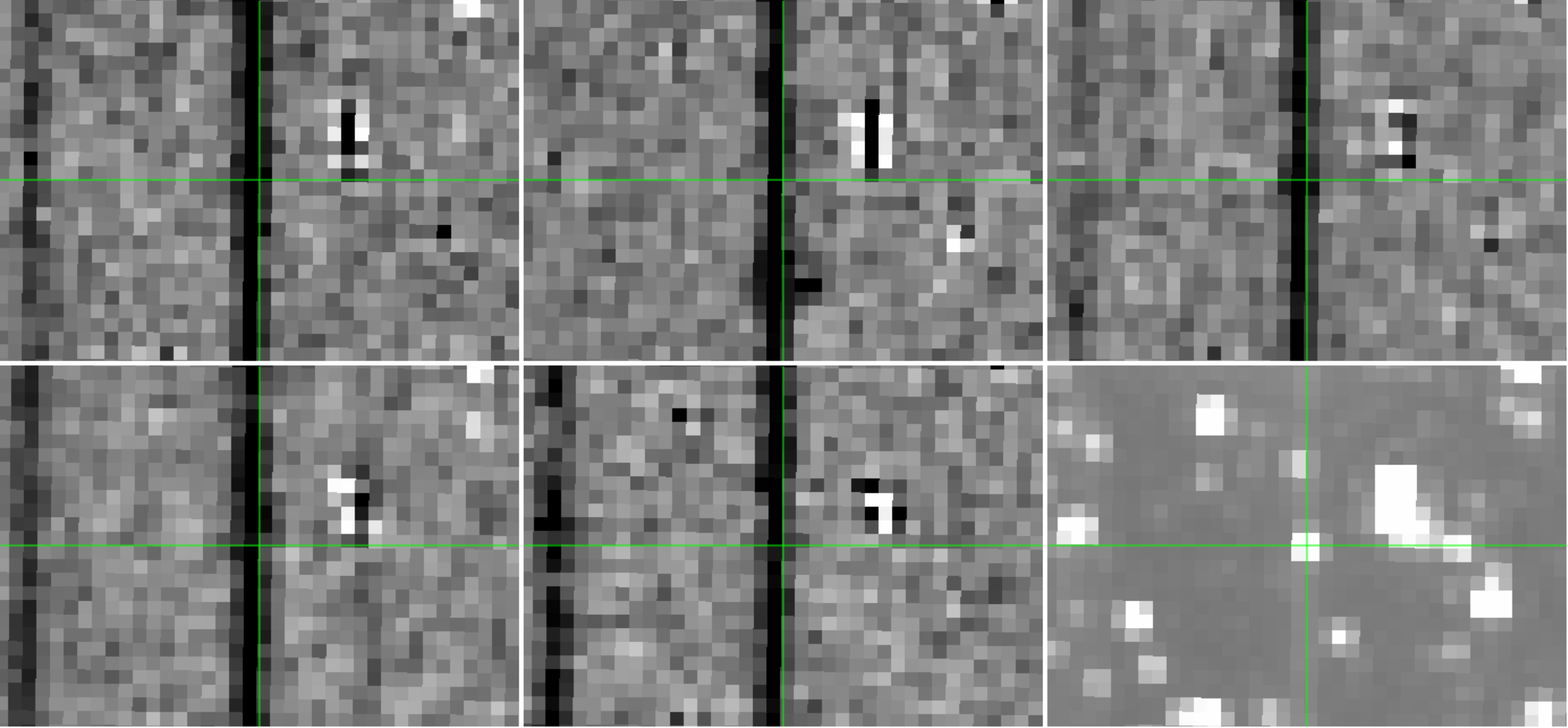}
	\includegraphics[width=0.5\textwidth]{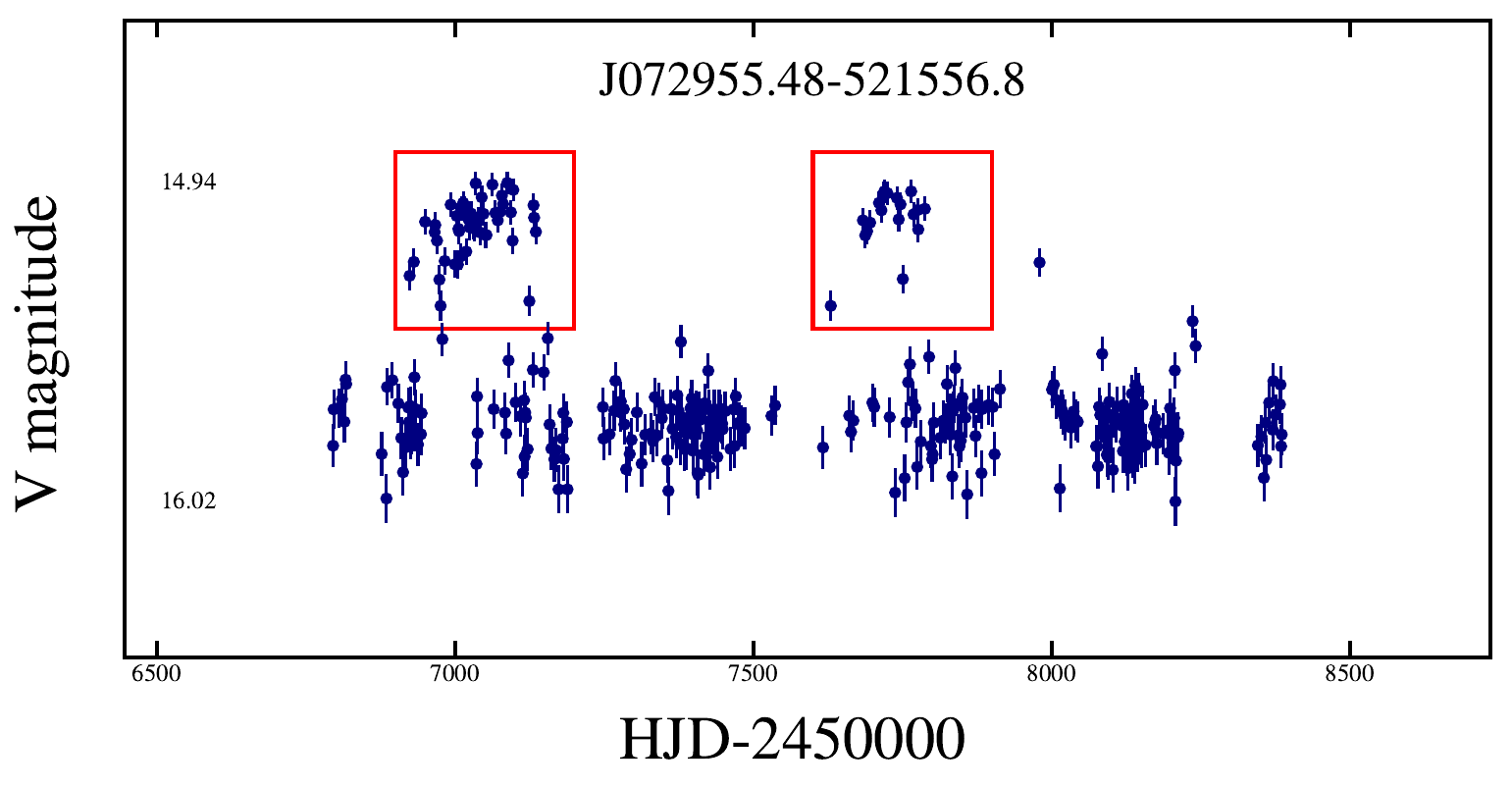}
    \caption{\textit{Top}: Examples of images that are affected by a malfunctioning shutter for the non-variable source J072955.48-521556.8, with the reference image shown in the bottom right, \textit{bottom}: The light curve for the source J072955.48-521556.8 with the affected epochs highlighted by the red boxes.}
    \label{fig:fig3}
\end{figure}

We also identified a systematic issue that affects the light curves of certain sources due to malfunctioning shutters. The shutters in the ASAS-SN cameras periodically start to fail. Degraded shutters do not close completely, and stray light from neighboring bright sources can impart trails on the images during readout (Figure \ref{fig:fig3}). This systematic is illustrated in the light curve for the non-variable source J072955.48-521556.8 (Figure \ref{fig:fig3}). We see that after the malfunctioning shutter is replaced, the light curve returns to normal, but after ${\sim}700$ days, the shutter begins to fail again. Some fraction of our ASAS-SN light curves will be affected by this systematic.

\section{Variability Analysis}

Here we describe the procedure we used to identify and characterize variables in the source list. We describe how we cross-matched the APASS sources to external catalogues in Section $\S3.1$. In Section $\S3.2$, we describe the procedure we took to identify candidate variable sources. In Section $\S3.3$, we discuss the application of the V2 random forest classifier model from \citet{2018arXiv180907329J} to classify these variables, and in Section $\S3.4$, we discuss the corrections done to mitigate the effects of blending on the candidate variables.

\subsection{Cross-matches to external catalogs}
We identify cross-matches to the APASS sources with Gaia DR2 \citep{2018arXiv180409365G} using the pre-computed cross matches from \citet{2018arXiv180809151M}. The sources were also cross-matched to the probabilistic distance estimates from \citet{2018AJ....156...58B}.
We also crossmatch the sources with 2MASS \citep{2006AJ....131.1163S} and AllWISE \citep{2013yCat.2328....0C,2010AJ....140.1868W} using a matching radius of 10\farcs0. We used \verb"TOPCAT" \citep{2005ASPC..347...29T} to cross-match the APASS sources with the 2MASS and AllWISE catalogs .

The Large Magellanic Cloud (LMC) lies within the TESS southern CVZ. We used Gaia DR2 \citep{2018A&A...616A..12G} to identify ${\sim}119,000$ sources from our source list that are LMC members. For sources in the LMC, we use a distance of $d=49.97$ kpc \citep{2013Natur.495...76P} in our variability classifier.
\subsection{Variability cutoffs}

There are numerous methods to identify variable sources from a sample of light curves. The most commonly used method involves correlating the variations observed in multiple bands to pick out `true' variables from the false positives \citep{1996PASP..108..851S}. The ASAS-SN observations used in this work are made with a single filter, which makes the use of multi-band variability statistics impossible. In this work, we used several methods, including periodogram statistics, light curve features and external photometry to identify variable sources.

\begin{figure*}
	\includegraphics[width=\textwidth]{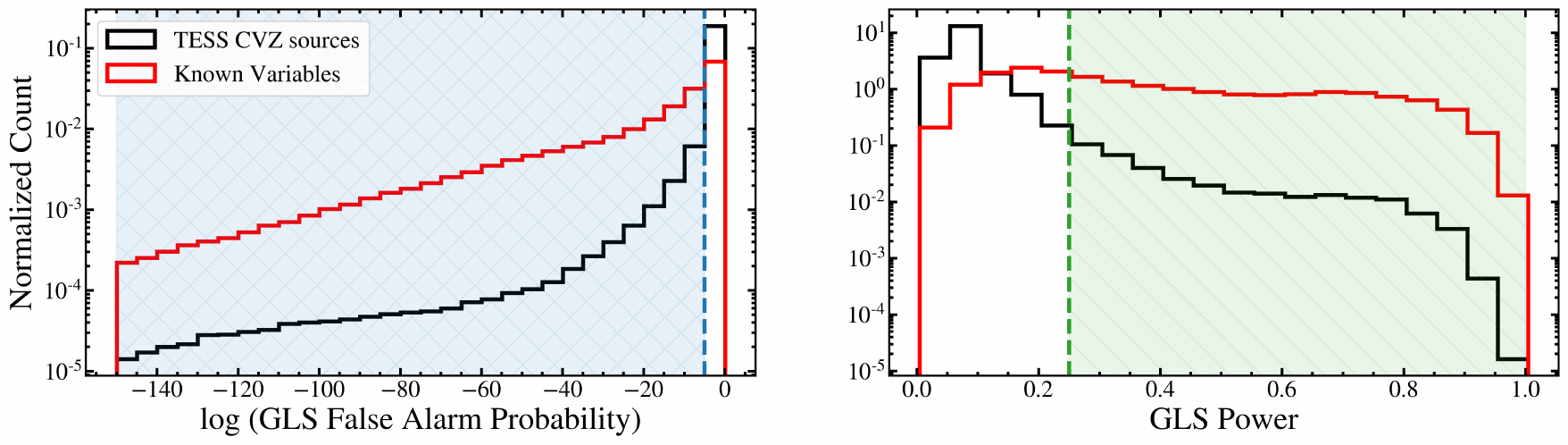}
    \caption{The distribution of the GLS false alarm probability (left), and the distribution of GLS power (right) for the ${\sim}1.3$M sources (black) and the set of known variables from \citet{2018arXiv180907329J} (red). Variable candidates had to lie in the shaded regions with $\log (\rm FAP)<-10$ and $\rm GLS\, Power>0.25$.}
    \label{fig:fig4}
\end{figure*}

We used the \verb"astropy" implementation of the Generalized Lomb-Scargle (GLS, \citealt{2009A&A...496..577Z,1982ApJ...263..835S}) periodogram to search for periodicity over the range $0.05\leq P \leq1000$ days for each of the ${\sim}1.3$M sources. We utilize the false alarm probability (FAP) and the power of the best GLS period as means of identifying significantly periodic sources. Sources with $\log (\rm FAP)<-10$ and $\rm GLS\, Power>0.25$ were selected for further analysis (Figure \ref{fig:fig4}).

\begin{figure*}
	\includegraphics[width=\textwidth]{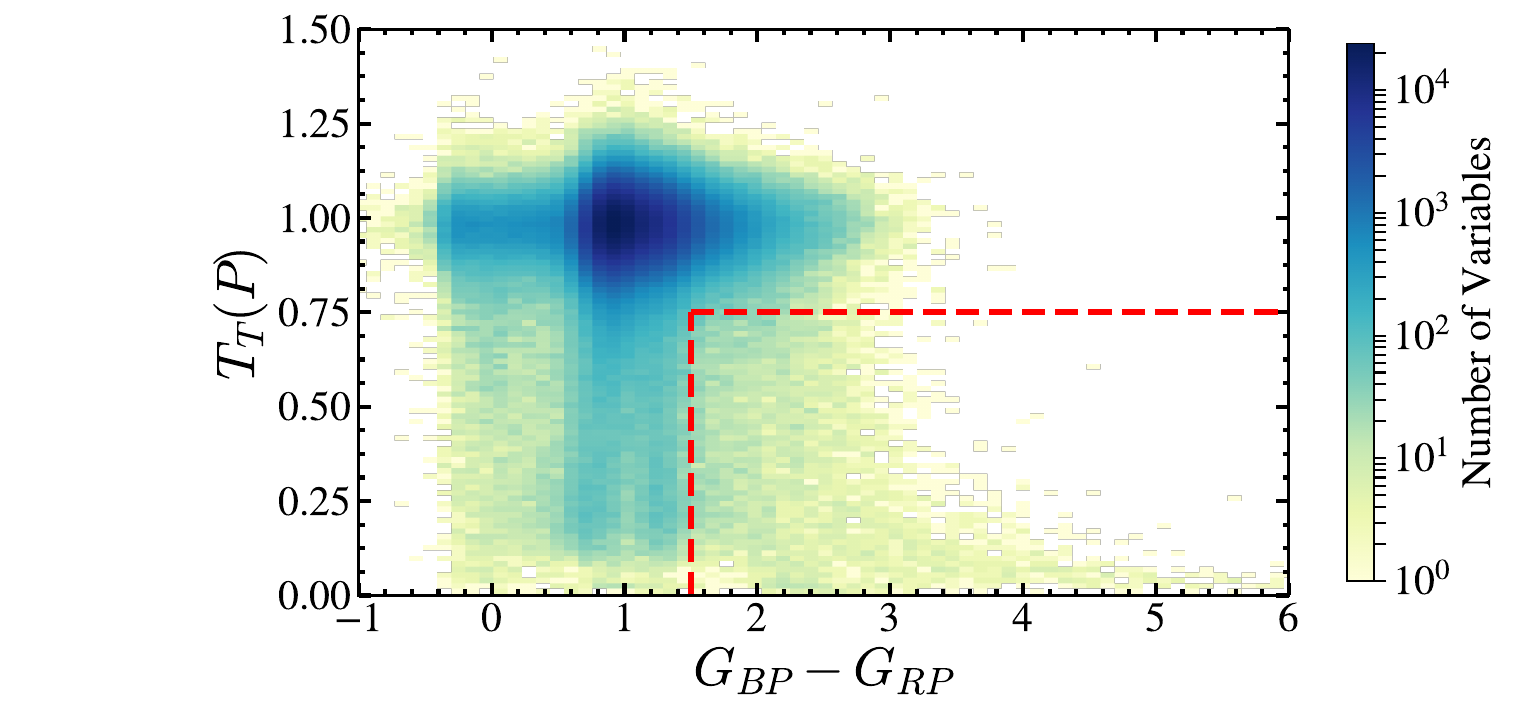}
    \caption{The distribution of $G_{BP}-G_{RP}$ with $T(t)$ for the ${\sim}1.3$M sources. The red shaded box encloses the sources that meet the criteria of $T(t)<0.75$ and $G_{BP}-G_{RP}>1.5$ mag used to identify red variable sources.}
    \label{fig:fig5}
\end{figure*}

We calculated the Lafler-Kinmann \citep{1965ApJS...11..216L,2002A&A...386..763C} string length statistic $T(t)$  on the temporal light curve using the definition
\begin{equation}
    T(t)=\frac{\sum_{i=1}^{\rm N} (m_{i+1}-m_i)^2}{\sum_{i=1}^{\rm N} (m_{i}-\overline m)^2}\times \frac{(N-1)}{2N}
	\label{eq:tt}
\end{equation} from \citet{2002A&A...386..763C}, where the $m_i$ are the magnitudes sorted temporally and $\overline m$ is the mean magnitude. We can also calculate this statistic sorting the light curve based on phase for a given period, which we will call $T(\phi|P)$. To identify red variables, we empirically isolate sources with $T(t)<0.75$ and the Gaia DR2 color $G_{BP}-G_{RP}>1.5$ mag (Figure \ref{fig:fig5}). Red variables typically have long periods that result in noticeable structure in their temporal light curves. This structure results in smaller values of $T(t)$ for the light curves of long period variables when compared to sources with short term variability (see \citealt{2018arXiv180907329J}).

\begin{figure*}
	\includegraphics[width=\textwidth]{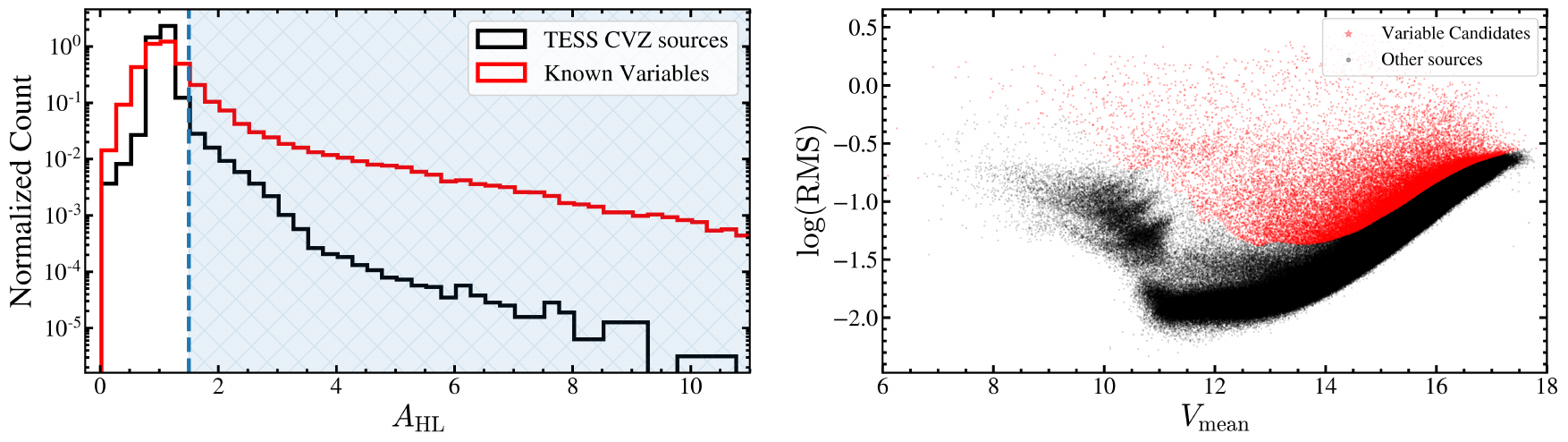}
    \caption{The distribution of $A_{\rm HL}$ (left), and the distribution of $\log{\rm RMS}$ with mean magnitude (right) for the ${\sim}1.3$M sources. The sources with $A_{\rm HL}>1.5$ are shaded in blue and the sources with $\log{\rm RMS}> 95^{\rm th}$ percentile in each magnitude bin are shaded in red. The distribution of $A_{\rm HL}$ for the set of known variables from \citet{2018arXiv180907329J} is plotted in red. }
    \label{fig:fig6}
\end{figure*}

We also compute the ratio of magnitudes brighter or fainter than average ($A_{\rm HL}$; \citealt{upsilon,2018arXiv180907329J}) for all the sources. We found in Paper II that eclipsing binaries have larger values of $A_{\rm HL}$ than most variables, so we flag sources with $A_{\rm HL}>1.5$ for further analysis. This variability cut is expected to improve the identification of detached eclipsing binaries. We also identify sources with a light curve RMS larger than the $95^{\rm th}$ percentile for the other stars in magnitude bins of $0.25$ mag (Figure \ref{fig:fig6}) in order to select sources with significant flux variations.

The variability cuts and the number of variable candidates isolated through each cut are summarized in Table \ref{tab:varcut}. The combination of these cuts help identify different variable sources and increase our completeness when compared to relying on just one or two parameters. Through these variability cutoffs, we identified ${\sim}60,000$ unique candidates. This amounts to ${\sim}5\%$ of the sources on the initial list. 

\begin{table*}
	\centering
	\caption{Summary of the variability selection cuts}
	\label{tab:varcut}
	\begin{tabular}{lrrr}
		\hline
		Variable(s) & Cut & Used to identify & Sources\\
		\hline
        GLS Power & $>0.25$ & Periodic Variables & 24257\\
        GLS $\log \, (\rm FAP)$ & $<-10$ & Periodic Variables & 39761\\ 
        \hline
        $G_{BP}-G_{RP}$ and $T(t)$ & $>1.5$ AND $<0.75$  & Red Variables & 7741\\ 
        \hline
        $A_{\rm HL}$ & $>1.5$ & Eclipsing Binaries & 20864\\
        \hline
        $\log \, (\rm RMS)$ & $>95^{\rm th}$ percentile & Large variations & 45486\\           
		\hline
	\end{tabular}

\end{table*}

\subsection{Variability Classification}
We derived periods for the ${\sim}60,000$ variable candidates following the procedure described in \citet{2018MNRAS.477.3145J,2018arXiv180907329J}. The \verb"astrobase" implementation \citep{astrob} of the Generalized Lomb-Scargle (GLS, \citealt{2009A&A...496..577Z,1982ApJ...263..835S}), the Multi-Harmonic Analysis Of Variance (MHAOV, \citealt{1996ApJ...460L.107S}), and the Box Least Squares (BLS, \citealt{2002A&A...391..369K}) periodograms were used to search for periodicity in these light curves. For each periodic source, we calculate the improvement in the Lafler-Kinmann string length statistic when phased with the best period ($T(\phi|P)$) when compared to the string length statistic calculated on the temporal light curve ($T(t)$),
\begin{equation}
    \delta=\frac{T(\phi|P)-T(t)}{T(t)}.
	\label{eq:tp}
\end{equation}

Following the period search, we use the variability classifier implemented in \citet{2018arXiv180907329J} to classify these variable candidates. We choose to visually review the classifications in order to improve our catalog. For the visual review, we select the sources with periods $P<40$ d if $\delta<0.05$ AND $T(t)<0.75$. The majority of the periodic variables within this period range should show significant improvements in the string length statistic when phased with a period. Sources with $P>40$ d are selected if they have $T(t)<1$. All irregular and aperiodic sources are also selected for visual review. In total, we visually reviewed  ${\sim}23,000$ variable candidates. During the process of visual review, we identified incorrect classifications ($3\%$) and periods ($4\%$) and corrected them. Sources with significant systematic and spurious variability were removed ($46\%$). We also changed the classifications of ${\sim}1300$ ($6\%$) sources to the generic variability type (`VAR'). At this stage, our list of variables consisted of ${\sim}12,300$ sources. This means that our initial candidate list had a false positive rate of ${\sim}80\%$. Part of this is that we were deliberately generous in our initial selection so that we can use the results to improve this aspect of the pipeline as we progress with our effort to identify variable sources over the full sky.

\subsection{Blending Corrections}
The large pixel scale of the ASAS-SN images (8\farcs0) and the FWHM (${\sim}$16\farcs0) results in blending towards crowded regions. The APASS catalog was constructed with images that have a significantly smaller pixel scale (2\farcs6), and as a result of this, multiple APASS sources can fall into a single ASAS-SN pixel. We do not correct for the contaminating light in the photometry of the blended sources, but we identify and correct blended variable groups in our catalog.

Since we extracted light curves for the positions of APASS sources, we can have two or more such sources inside a single ASAS-SN resolution element. If we select the sources with another APASS neighbor within 30\farcs0, we find that 559 of the ${\sim}12,300$ variables had a neighbor within 30\farcs0. We compute the flux variability amplitudes for these sources using a random forest regression model \citep{2018arXiv180907329J}. The majority of the variable groups consisted of two sources, with a few groups consisting of up to three sources. For each variable group, we consider the source with the largest flux variability as the `true' variable, and remove the other overlapping sources from the final list. Following this treatment, our list of variables consisted of ${\sim}11,700$ sources.

\section{Results}

The complete catalog of ${\sim}11,700$ variables is available at the ASAS-SN Variable Stars Database (\href{https://asas-sn.osu.edu/variables}{https://asas-sn.osu.edu/variables}) along with the V-band light curves for each source. Table \ref{tab:var} lists the number of sources of each variability type in the catalog. 

\begin{table*}
	\centering
	\caption{Variables by type}
	\label{tab:var}
\begin{tabular}{llrrr}
		\hline
		VSX Type & Description & LMC & Not LMC & New discoveries\\
		\hline
CWA   & W Virginis type variables with $P>8$ d & 10 & -- & 1\\
CWB   & W Virginis type variables with $P<8$ d & 4 & 8 & 2\\
DCEP  & Fundamental mode Classical Cepheids& 577 & 12 & 12\\
DCEPS & First overtone Cepheids & 280 & 9 & 10\\
DSCT  & $\delta$ Scuti variables & -- & 44 & 42\\
EA    & Detached Algol-type binaries & 103 & 1109 & 865\\
EB    & $\beta$ Lyrae-type binaries & 129 & 523 & 353\\
EW    & W Ursae Majoris type binaries & 97 & 1961 & 1228\\
ELL   & Ellipsoidal Variables & 1 & 11 & 12\\
HADS  & High amplitude $\delta$ Scuti variables & -- & 102 & 61\\
ROT   & Rotational variables & -- & 881& 813\\
RRAB  & RR Lyrae variables (Type ab) & 42 & 444& 44\\
RRC   & First Overtone RR Lyrae variables & 4 & 443& 217\\
RRD   & Double Mode RR Lyrae variables & -- & 4& 3\\
RVA   & RV Tauri variables (Subtype A) & 9 & 1& 1\\
SR    & Semi-regular variables & 1017 & 1487& 1340\\
\hline
L     & Irregular variables & 298 & 137& 262\\
GCAS  & $\gamma$ Cassiopeiae variables & 191 & 49& 128\\
YSO   & Young stellar objects & 2 & 8& 6\\
\hline
ROT:   & Uncertain rotational variables & -- & 186& 174\\
DSCT:  & Uncertain $\delta$ Scuti variables & 4 & 63& 48\\
GCAS:  & Uncertain $\gamma$ Cassiopeiae variables & 10 & --& 3\\
VAR  & Generic variables & 40 & 1390 & 1385\\
\hline
\end{tabular}
\end{table*}

\begin{figure*}
	\includegraphics[width=\textwidth]{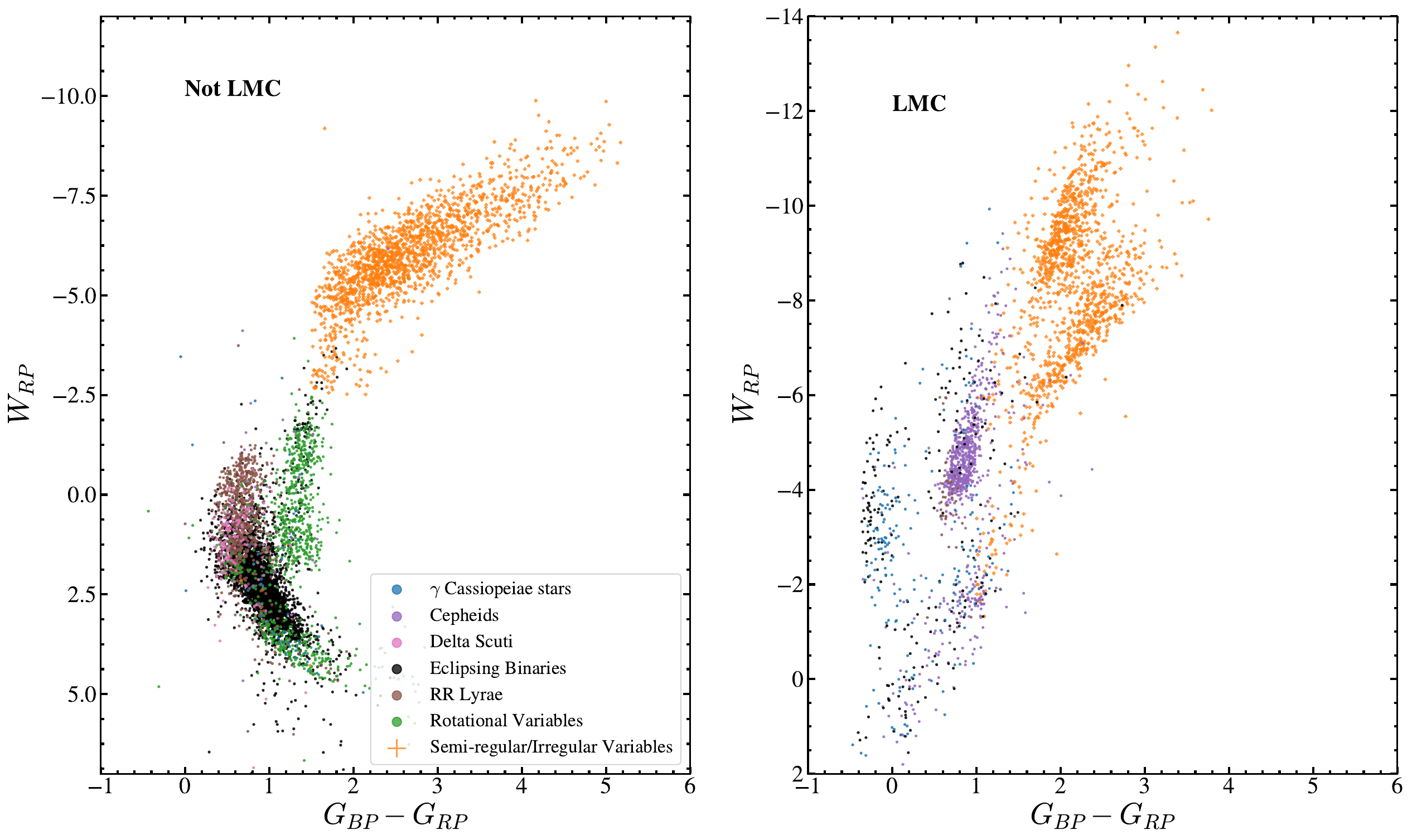}
    \caption{The Wesenheit $W_{RP}$ vs. $G_{BP}-G_{RP}$ color-magnitude diagram for the variables, outside the LMC (left), and in the LMC (right).}
    \label{fig:fig7}
\end{figure*}

In paper II, we used the reddening-free Wesenheit magnitudes \citep{1982ApJ...253..575M,2018arXiv180803659L} \begin{equation}
    W_{RP}=M_{\rm G_{RP}}-1.3(G_{BP}-G_{RP}) \,, 
	\label{eq:wrp}
\end{equation} 
and
\begin{equation}
    W_{JK}=M_{\rm K_s}-0.686(J-K_s) \,
	\label{eq:wk}
\end{equation}
for variability classification. The Wesenheit $W_{RP}$ vs. $G_{BP}-G_{RP}$ color-magnitude diagram for all the variables is shown in Figure \ref{fig:fig7}. We have sorted the variables into groups to highlight the different classes of variable sources. Owing to the magnitude limit of ASAS-SN, we are only able to detect sufficiently bright sources ($V\lesssim16$ mag, $W_{RP}\lesssim1.5$ mag) in the LMC. This is evident in the Wesenheit color-magnitude diagram for the LMC sources. For the sources outside the LMC, we probe a wider range of magnitudes ($W_{RP}\lesssim6$ mag), including RR Lyrae, rotational variables and $\delta$ Scuti variables. 

\begin{figure*}
	\includegraphics[width=\textwidth]{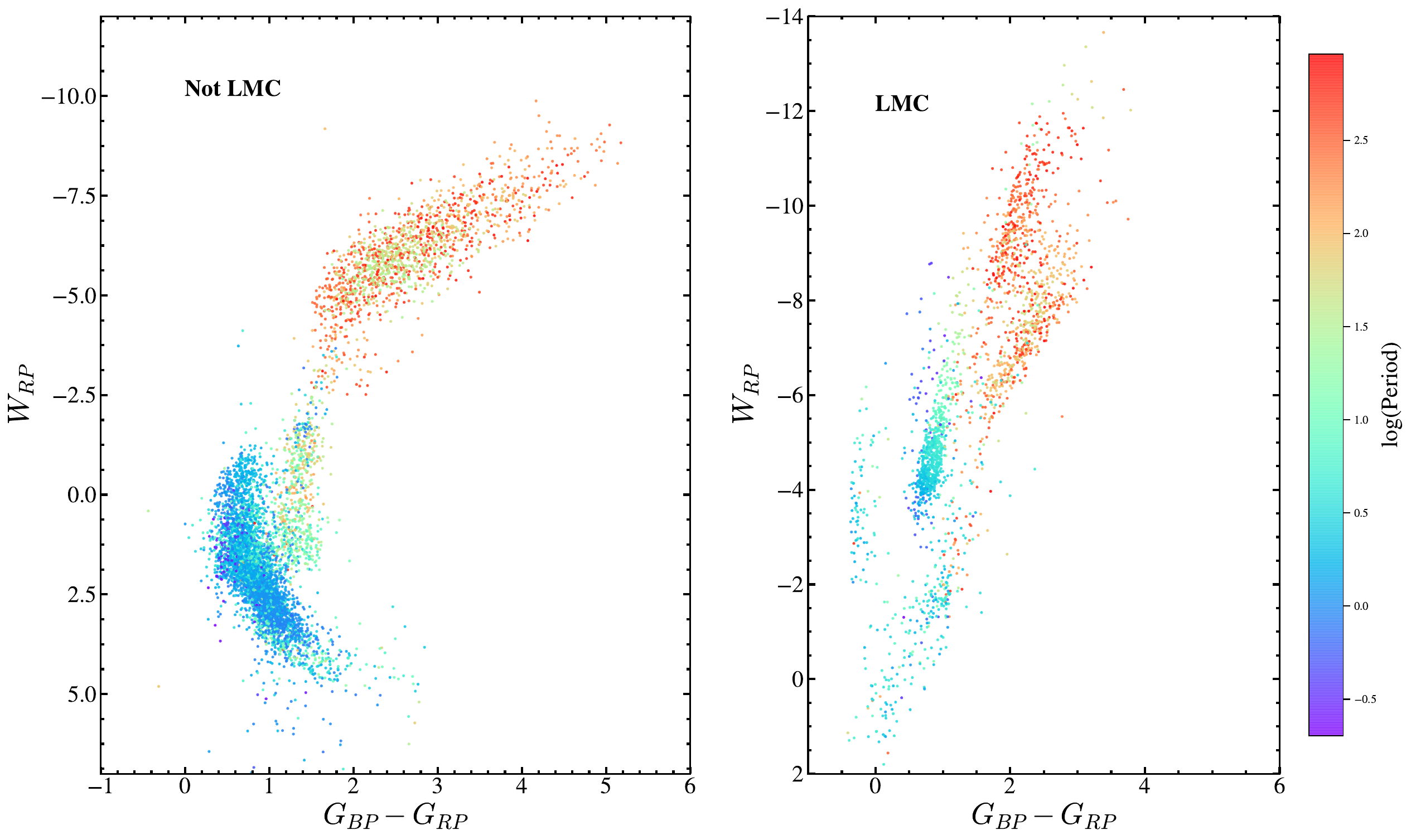}
    \caption{The Wesenheit $W_{RP}$ vs. $G_{BP}-G_{RP}$ color-magnitude diagram for the periodic variables, outside the LMC (left), and in the LMC (right). The points are colored by the period.}
    \label{fig:fig8}
\end{figure*}

We have also plotted the Wesenheit $W_{RP}$ vs. $G_{BP}-G_{RP}$ color-magnitude diagram for all the periodic variables in Figure \ref{fig:fig8}, with the points colored according to the period. This essentially highlights the large dynamic range in period probed by the ASAS-SN light curves.

We note that the identification of Mira variables in the LMC is hindered by blending. Due to blended light, the observed ASAS-SN amplitudes of these sources fall below the amplitude threshold of $V>2$ mag that is used to define a Mira variable in our pipeline. In reality, some fraction of the LMC semi-regular variables in this catalog are actually Mira variables. From the sample of known semi-regular variables in the LMC, we estimate that the fraction of Mira variables classified as semi-regular variables is ${\sim}23\%$.

\begin{figure*}
	\includegraphics[width=\textwidth]{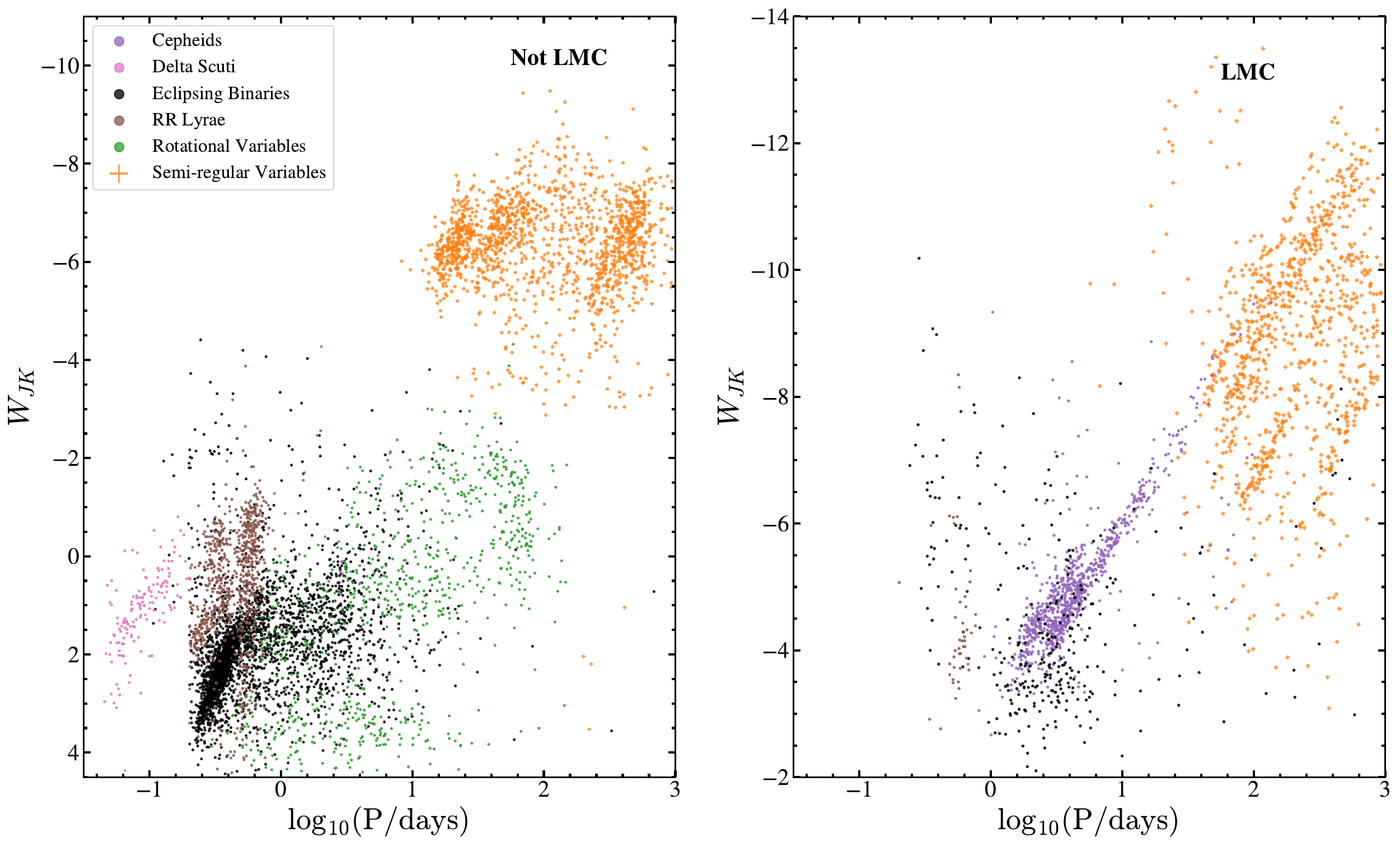}
    \caption{The Wesenheit $W_{JK}$ PLR diagram for the periodic variables, outside the LMC (left), and in the LMC (right). The points are colored as in Figure \ref{fig:fig7}.}
    \label{fig:fig9}
\end{figure*}

Using the same color scheme, the combined Wesenheit $W_{JK}$ PLR diagram for the periodic variables is shown in Figure \ref{fig:fig9}. The PLR sequences for the Cepheids and semi-regular variables in the LMC are well defined \citep{2005AcA....55..331S}. Most of the eclipsing binaries identified in the LMC are either detached or semi-detached systems, and do not follow the well defined PLR for contact binaries that is observed in the PLR diagram for the sources outside the LMC.

\begin{figure*}
	\includegraphics[width=\textwidth]{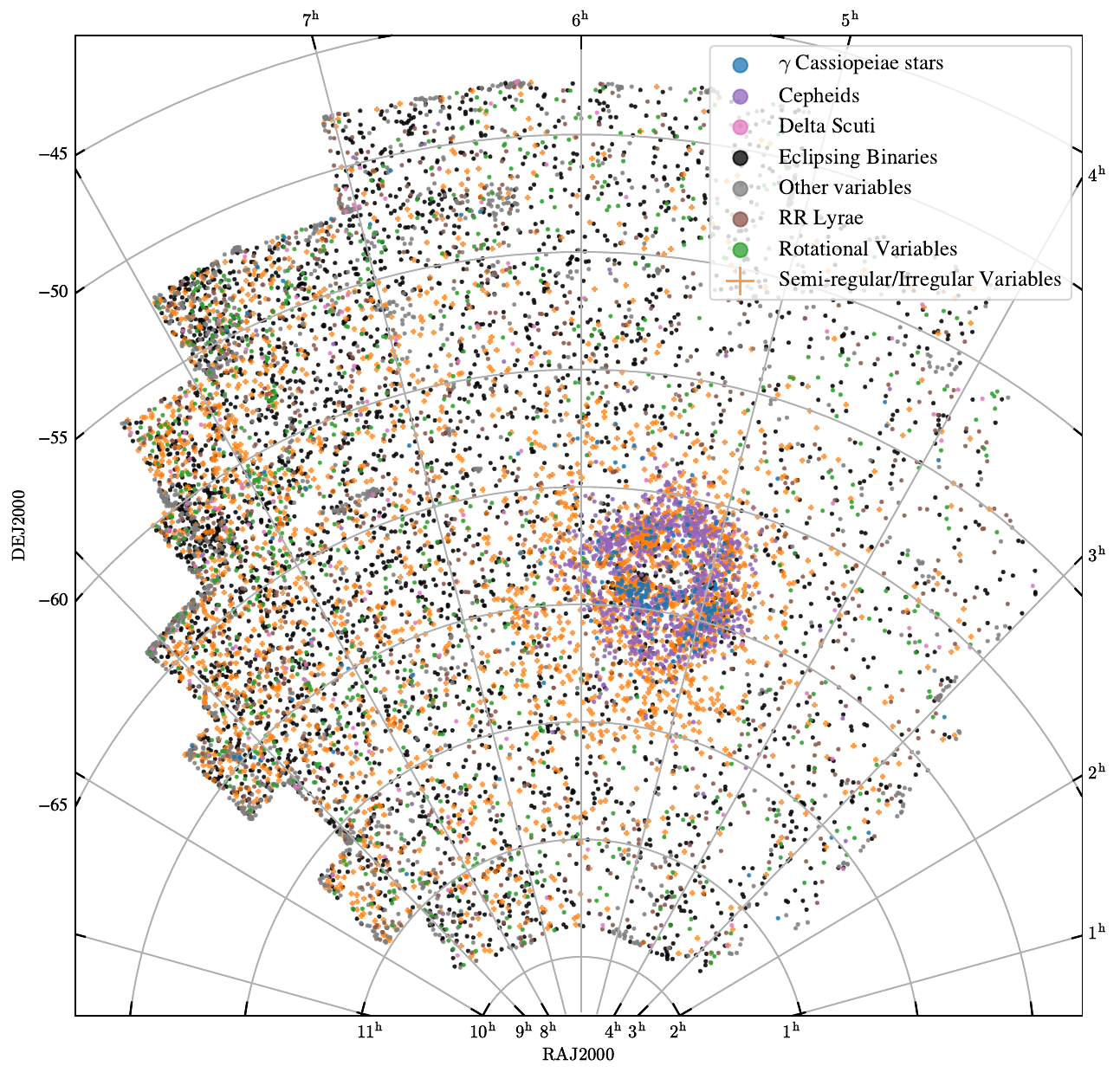}
    \caption{The sky distribution of the variables in equatorial coordinates. The points are colored as in Figure \ref{fig:fig7}.}
    \label{fig:fig10}
\end{figure*}

We also show the sky distribution of the variables identified in this work in Figure \ref{fig:fig10}. We see that the distribution of eclipsing binaries and rotational variables (black points) is random, but the distribution of Cepheids is strongly clustered towards the LMC as is expected. We also note the clustering of semi-regular/irregular variables (red giants) towards the LMC and the Galactic disk.

We matched our list of variables to the VSX \citep{2006SASS...25...47W} catalog available in October 2018, with a matching radius of 16\farcs0 to identify previously discovered variables. The variables discovered by the All-Sky Automated Survey (ASAS; \citealt{2002AcA....52..397P}) and the Catalina Real-Time Transient Survey (CRTS; \citealt{2014ApJS..213....9D}) are included in the VSX database. We also match our variables to the catalogs of variable stars discovered by ASAS-SN \citep{2018MNRAS.477.3145J}, the catalogs of variable stars in the Magellanic clouds and the Galactic bulge from the Optical Gravitational Lensing Experiment (OGLE; \citealt{2003AcA....53..291U,2016AcA....66..421P,2016AcA....66..405S}, and references therein), the Gaia DR2 catalog of variables \citep{2018arXiv180409365G,2018arXiv180409373H,gdr2var}, the catalog of variables from the Asteroid Terrestrial-impact Last Alert System (ATLAS; \citealt{2018PASP..130f4505T,2018arXiv180402132H}), the catalog of KELT variables \citep{2018AJ....155...39O} and the variables from MACHO \citep{1997ApJ...486..697A}. Of the ${\sim}11,700$ variables identified in this work, ${\sim}4,700$ were previously discovered by other surveys, as also listed in Table \ref{tab:var}. The majority of these known variables consist of eclipsing binaries (${\sim}31\%$), irregular/semi-regular variables (${\sim}28\%$) and Cepheids (${\sim}19\%$). 

This leaves ${\sim}7,000$ new variables. Most of the new discoveries are eclipsing binaries (${\sim}35\%$) and irregular/semi-regular variables (${\sim}23\%$). We also discovered 128 new GCAS variables, 81 of which are located in the LMC. GCAS variables are rapidly rotating, early-type irregular variable stars (typically Be stars) with mass outflow from their equatorial regions. Example light curves for the newly identified GCAS variables are shown in Figure \ref{fig:fig12}. 

\begin{figure*}
	\includegraphics[width=\textwidth]{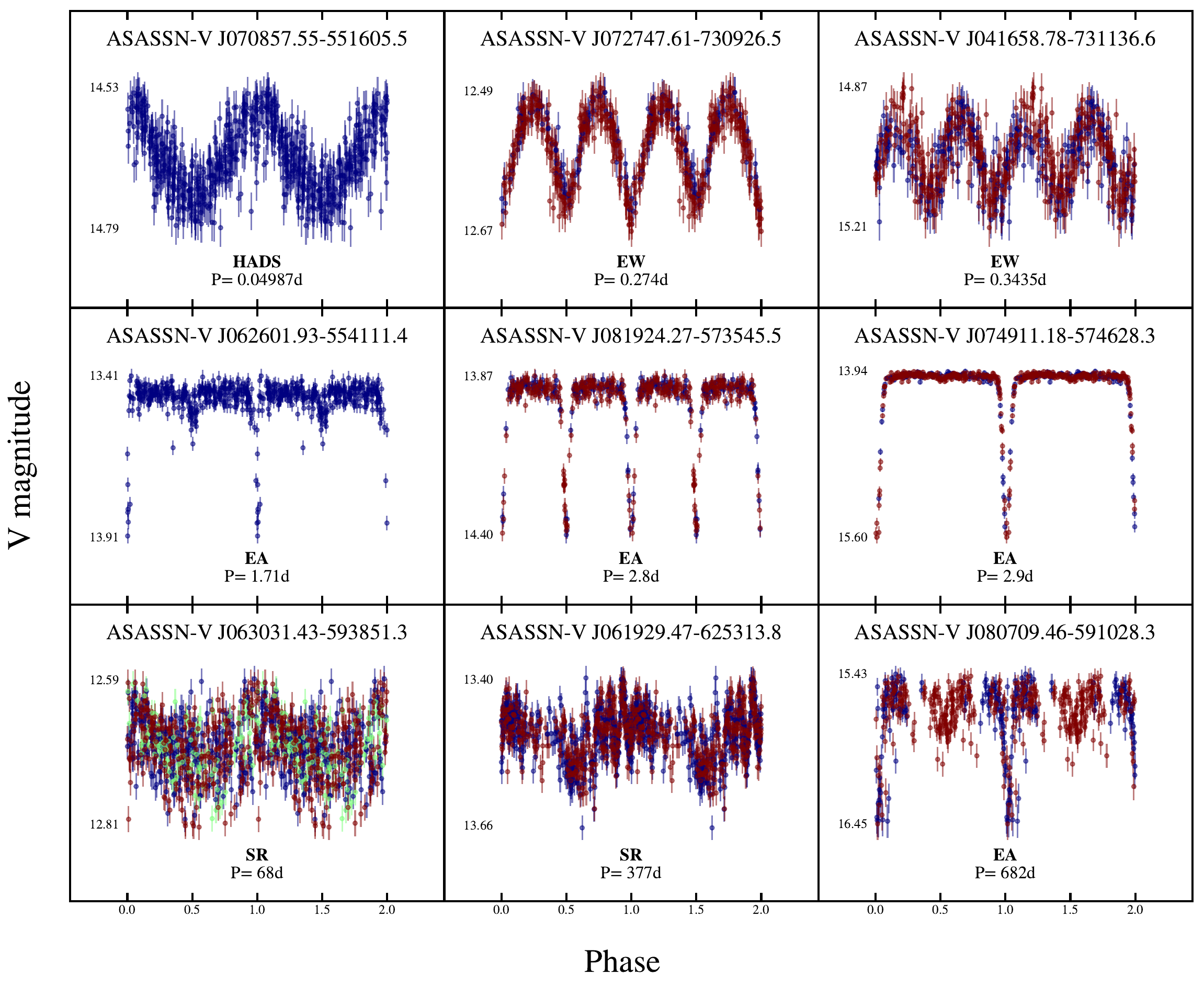}
    \caption{Phased light curves for examples of the newly discovered periodic variables. The light curves are scaled by their minimum and maximum V-band magnitudes. Different colored points correspond to data from the different ASAS-SN cameras. The different variability types are defined in Table \ref{tab:var}.}
    \label{fig:fig11}
\end{figure*}

\begin{figure*}
	\includegraphics[width=\textwidth]{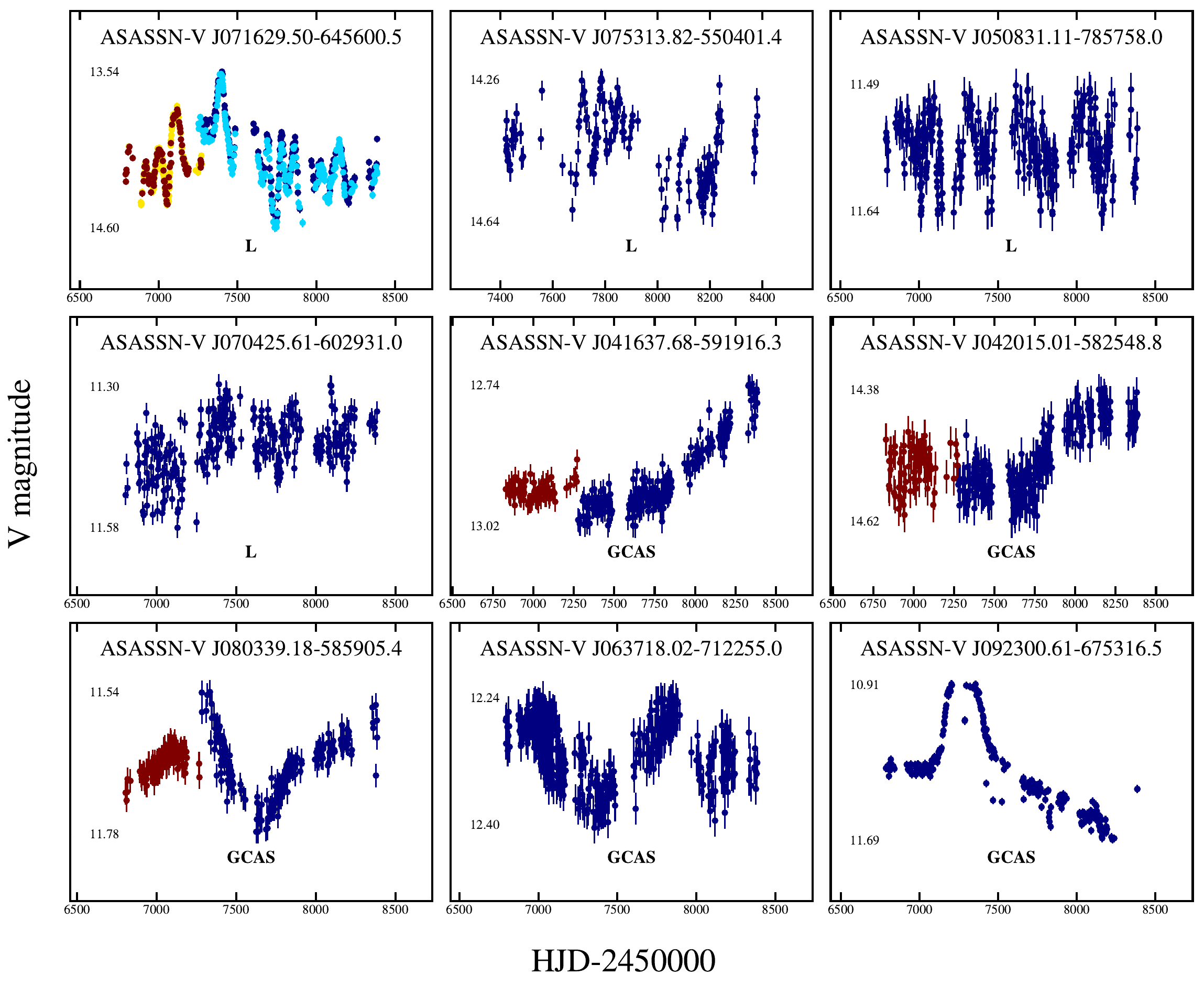}
    \caption{Light curves for examples of the newly discovered irregular variables. The format is the same as for Fig. \ref{fig:fig11}}
    \label{fig:fig12}
\end{figure*}

We also discovered another long period detached eclipsing binary (ASASSN-V J080709.46$-$591028.3) in this work. ASASSN-V J080709.46$-$591028.3 is located away from the LMC and has an orbital period of $P_{\rm orb}{\sim}682$ d (${\sim}1.9$ yr). Its light curve shows evidence of both a primary and secondary eclipse with a primary eclipse depth of ${\sim}1$ mag (Figure \ref{fig:fig11}).

A non-negligible fraction of these new discoveries were classified as generic variables (${\sim}20\%$). These are mostly low amplitude or faint sources. Example light curves are shown for a subset of the newly discovered periodic (irregular) variables in Figure \ref{fig:fig9} (\ref{fig:fig10}).

\section{Conclusions}

We systematically searched for variable sources in a ${\sim}1000 \, \rm deg^2$ region surrounding the Southern ecliptic pole. This region is coincident with the Southern continuous viewing zone for the TESS satellite and thus, a large majority of these sources will have excellent TESS light curves.

Through our search, we identified ${\sim}11,700$ variable sources, of which ${\sim}7,000$ are new discoveries. Variable sources identified in the LMC largely consist of luminous variables, including Cepheids, GCAS variables (Be stars) and red giants. We identify a broader sample of variables outside the LMC, including RR Lyrae, eclipsing binaries, rotational variables and $\delta$ Scuti variables. 

We have developed a user friendly interface to retrieve pre-computed ASAS-SN V-band light curves for APASS sources. The V-band light curves of all the ${\sim}1.3$M sources studied in this work are available online at the ASAS-SN Photometry Database (\url{https://asas-sn.osu.edu/photometry}). To highlight the possible blended sources, a flag is assigned to each source if the distance to the nearest APASS neighbor is <16\farcs0. The new variable sources have also been added to the ASAS-SN variable stars database (\url{https://asas-sn.osu.edu/variables}).

As part of our ongoing effort to systematically analyze all the ${\sim}50$ million $V<17$ mag APASS sources for variability, we will gradually update this database with the light curves for the sources across the remainder of the sky over the course of 2019. This work provides long baseline V-band light curves for a large fraction of the sources in the TESS southern CVZ and is a useful supplement to the short baseline TESS light curves that possess better photometric precision. 

\section*{Acknowledgements}

We thank the referee for their useful comments. We thank the Las Cumbres Observatory and its staff for its
continuing support of the ASAS-SN project. We also thank the Ohio State University College of Arts and Sciences Technology Services for helping us set up and maintain the ASAS-SN variable stars and photometry databases.

ASAS-SN is supported by the Gordon and Betty Moore
Foundation through grant GBMF5490 to the Ohio State
University and NSF grant AST-1515927. Development of
ASAS-SN has been supported by NSF grant AST-0908816,
the Mt. Cuba Astronomical Foundation, the Center for Cos-
mology and AstroParticle Physics at the Ohio State Univer-
sity, the Chinese Academy of Sciences South America Center
for Astronomy (CAS- SACA), the Villum Foundation, and
George Skestos. 

This work is supported in part by Scialog Scholar grant 24216 from the Research Corporation. TAT acknowledges support from a Simons Foundation Fellowship and from an IBM Einstein Fellowship from the Institute for Advanced Study, Princeton. Support for JLP is provided in part by the Ministry of Economy, Development, and Tourism's Millennium Science Initiative through grant IC120009, awarded to The Millennium Institute of Astrophysics, MAS. SD acknowledges Project 11573003 supported by NSFC. Support for MP and OP has been provided by the PRIMUS/SCI/17 award from Charles University. This work was partly supported by NSFC 11721303.

This work has made use of data from the European Space Agency (ESA)
mission {\it Gaia} (\url{https://www.cosmos.esa.int/gaia}), processed by
the {\it Gaia} Data Processing and Analysis Consortium. This publication makes 
use of data products from the Two Micron All Sky Survey, as well as
data products from the Wide-field Infrared Survey Explorer.
This research was also made possible through the use of the AAVSO Photometric 
All-Sky Survey (APASS), funded by the Robert Martin Ayers Sciences Fund. 

This research has made use of the VizieR catalogue access tool, CDS, Strasbourg, France. 
This research made ualso se of Astropy, a community-developed core Python package for 
Astronomy (Astropy Collaboration, 2013).










\bsp	
\label{lastpage}
\end{document}